\documentclass{emulateapj}

\begin{document}

\title{Dynamical Masses in Luminous Infrared Galaxies\altaffilmark{1}}
\author{J. L. Hinz and G. H. Rieke}
\affil{Steward Observatory, University of Arizona, 933 N. Cherry Ave.,  Tucson, AZ  85721 
\\email: jhinz, grieke@as.arizona.edu}

\altaffiltext{1}{Observations reported here were obtained at the MMT Observatory, a joint facility of the University of Arizona and the Smithsonian Institution.}

\begin{abstract}

We have studied the dynamics and masses of a sample of ten
nearby luminous and ultraluminous infrared galaxies (LIRGS and ULIRGs),
using 2.3\,$\micron$ 
$^{12}$CO absorption line 
spectroscopy and near-infrared $H$- and $K_s$-band imaging.  By combining velocity
dispersions derived from the spectroscopy, disk scale-lengths
obtained from the imaging, and a set of likely model density profiles, 
we calculate dynamical masses for each LIRG. For the majority of
the sample, it is difficult to reconcile our mass estimates with 
the large amounts of gas derived from millimeter observations and
from a standard conversion between $^{12}$CO emission and H$_2$ mass.

Our results imply that LIRGs
do not have huge amounts of molecular gas ($10^{10}-10^{11}$ 
M$_{\odot}$) at their centers, and support previous indications 
that the standard conversion of $^{12}$CO 
to H$_2$ probably overestimates the gas masses and cannot be used in
these environments.  This in turn suggests much more modest levels of 
extinction in the near-infrared for LIRGs than previously predicted 
(A$_V \sim$\,10-20 versus A$_V \sim$\,100-1000).  The lower gas mass estimates
indicated by our observations imply that the star formation efficiency in these 
systems is very high and is triggered by cloud-cloud collisions, shocks, and 
winds rather than by gravitational instabilities in circumnuclear gas disks.

\end{abstract}


\section{Introduction} \label{sec:introduction}

Luminous and ultraluminous infrared galaxies (LIRGs and ULIRGs) 
have extremely high bolometric
luminosities and have spectral energy distributions that are dominated
by infrared light (8-1000\,$\micron$ luminosities 
$\geq$ 10$^{11}\,L_{\odot}$ for LIRGs and 
$\geq$ 10$^{12}\,L_{\odot}$ for ULIRGs).  
Their morphologies are generally disturbed, sometimes
with tidal tails or two discernable nuclei, serving as strong
evidence that they are the products of mergers and interactions (e.g.,
Farrah et al. 2001; Bushouse et al. 2002; Veilleux et al. 2002).  Their
infrared power is thought to be from intense 
star formation triggered by the merging process, supplemented by
the outputs of active
galactic nuclei (AGNs). It has been suggested that such
systems go through not only a luminous starburst phase, but later evolve
into QSOs (Sanders et al. 1988).

LIRGs are often the result of the collision of two gas-rich disk 
galaxies and have been suggested to be the precursors to elliptical galaxies 
(Toomre \& Toomre 1972; Schweizer 1986; Hernquist 1992, 1993).  This
hypothesis can be tested using the fundamental plane, a three-parameter
relation among effective radius, velocity dispersion and mean
surface brightness for early-type galaxies (e.g., Djorgovski \& Davis 1987; 
Dressler et al. 1987).  It has been shown (Genzel et
al. 2001; Tacconi et al. 2002)
that most LIRGs are remarkably close to the
fundamental plane of dynamically hot galaxies, particularly on the 
less ``evolution-sensitive'' effective radius vs. velocity dispersion projection.
Only LIRGs with the smallest effective radii are offset from the plane.
The placement of LIRGs within the fundamental plane matches the locations
of intermediate-mass, disky (rotating) ellipticals.  The closeness of LIRGs
to ellipticals on the fundamental plane is somewhat surprising, but is
usually interpreted as a ``conspiracy'' of stellar evolution and 
extinction -- the surface brightening of the new stellar population
is obscured by heavy dust extinction to keep the LIRGs on the plane.
However, the dynamical similarities of the LIRGs and intermediate-mass
ellipticals are robust and do not depend on the photometric properties.
Therefore, LIRGs are 
likely an important link to understanding the formation and evolution
of early-type galaxies from gas-rich spirals.  This link is now
being strengthened by studies of the mid- and far-infrared properties of LIRGs and
ULIRGs at high redshifts (Charmandaris et al. 2004; Egami et al. 2004;
Le Floc'h et al. 2004), by ground-based integral field spectroscopic
studies investigating the dynamics of low redshift systems and their link
to deriving dynamical masses for LIRGs at higher redshifts 
(Colina et al. 2005), and by spectroscopic expansions on Genzel et al. (2001)
and Tacconi et al. (2002) studying dynamical evolution (Dasyra et al. 2006).

In the conventional view of LIRGs, one by-product of the merger of two
gas-rich disks is that huge masses of molecular gas will settle into
their nuclei.
Interferometric observations of LIRGs at millimeter wavelengths 
indicate large concentrations of molecular gas at their centers.
Calculations to derive gas masses using maps of the $^{12}$CO 
J$=1\rightarrow0$ transition sometimes 
adopt a standard conversion factor between the $^{12}$CO J\,$=1\rightarrow0$
luminosity and the molecular hydrogen (H$_2$) mass.  The conversion was
originally derived from observations of $^{12}$CO to H$_2$ gas in giant 
molecular clouds in our own galaxy (Scoville \& Good 1989) and takes the form

\begin{center}
\begin{math}
M_g = 1.2 \times 10^4\ F_{CO} D^2_{Mpc} (1+z)^{-1}\ M_{\odot}
\end{math}
\end{center}

\noindent where $M_g$ is the gas mass ($M$(H$_2$ + He)), 
$F_{CO}$ is the spatial- and velocity-integrated
CO flux in Jy km s$^{-1}$, 
$D_{Mpc}$ is the distance to the galaxy in Mpc, and $z$ is the redshift.
The standard conversion factor is equivalent to 
$N(H_2)/I_{CO}=2.24 \times 10^{20}$ cm$^{-2}$ (K km s$^{-1})^{-1}$.
This conversion produces molecular gas masses in the range
$10^{10}-10^{11} M_{\odot}$ (Bryant \& Scoville 1999), indicating 
extinctions as large as A$_V\sim100-1000$ in the optical (see, e.g.,
Narayanan et al. 2005 for recent alternative conversion factors).
If true, this would mean that even observations taken in the near-infrared
would barely penetrate the surface of most LIRGs and that much of their
star formation activity would remain hidden from view.

However, there are indications that the standard conversion
between CO flux and H$_2$ mass may not be applicable in the environment in LIRGs.  
Both theory and observations
show that the state of gas in the centers of LIRGs is quite
different from that in local giant molecular clouds.  Maloney \& Black (1988)
argued that the high densities and temperatures of the molecular gas in
dense starbursts would result in the standard ratio significantly 
overestimating the total mass.  They stated that, while the conversion does
seem to be constant over most of the Milky Way, the underlying reason for this
is not well understood, and should not be assumed for other galaxies. 
Additionally, a CO-to-H$_2$ ratio that varies across the Milky
Way presents a solution to the apparent observed discrepancy between the 
radial gradient in the diffuse Galactic $\gamma$-ray and the distribution 
of supernova remnants (Strong et al. 2004).  There is also evidence for 
variances in the standard conversion between galaxies, where arguments
for a strong dependence on metallicity as traced by the oxygen abundance
[O]/[H] have been presented (Israel 
1997; 2000; Barone et al. 2000; Boselli et al. 2002).

Solomon et al. (1997) and Downes \& Solomon (1998) concluded that the
standard factor may be high by a factor of five based on a comparison of
dynamical and gas masses derived from far-infrared observations.  Lisenfeld,
Isaak, \& Hills (2000) also suggested, using submillimeter measurements of
cold dust emission, that the molecular gas mass deduced from CO observations
may be 2-3 times higher than would be deduced from thermal emission.

Other works
have challenged the molecular gas masses calculated from the standard
conversion by comparing those values to measurements of the total dynamical
masses.  
Shier, Rieke, \& Rieke (1994) obtained dynamical masses for three
LIRGs based on velocity
dispersions from near-infrared spectra and scale-lengths derived from
near-infrared light profiles.  They found that in all three galaxies
the total dynamical masses of the cores were less than the claimed H$_2$
masses.  They concluded that, despite large errorbars on the dynamical
masses, the standard conversion factor must be wrong for at least some
luminous infrared galaxies.  Alonso-Herrero et al. (2001) calculated
a dynamical mass for NGC 1614 derived from Br\,$\gamma$ spectroscopy (Puxley
\& Brand 1999) and their knowledge of the morphology and size of a ring of
star formation revealed in {\it HST}/NICMOS Pa\,$\alpha$ images.
They found that the total dynamical mass of the starburst region is nearly four
times smaller than the mass of molecular gas estimated from the standard 
ratio of $^{12}$CO ($1\rightarrow0$) to H$_2$. 

The difference in estimated
masses is important for understanding the way in which
stars are formed in these galaxies.  For instance, if the large surface
densities of gas in the circumnuclear disks predicted by the standard ratio
are correct, then star formation occurs through instabilities in the disk 
(Taniguchi \& Ohyama 1998).  However, if lower mass estimates are valid, then 
star formation
through cloud-cloud collision, shocks, and winds (Scoville, Sanders \& 
Clemens 1986) is predicted.
Additionally, using the standard conversion suggests a low star formation
efficiency, whereas lower mass estimates yield high star formation efficiency. 

To understand more fully star formation and extinction in LIRGs, we
determine dynamical masses for a sample of ten galaxies. 
We describe the observations and data reduction in \S \ref{sec:observations}, 
discuss the analysis and modeling in \S \ref{sec:analysis}, compare the
molecular masses to dynamical masses for the sample in \S \ref{sec:masses},
and summarize the results in \S \ref{sec:summary}. 

\section{Observations and Data Reduction} \label{sec:observations}

Ten luminous and ultraluminous infrared galaxies were chosen for this study.  
These galaxies
have $K$-band magnitudes in the range $\sim$\,12-13.5.  All of the LIRGs
are nearby ($v <$ 17,000\,km s$^{-1}$), and many have millimeter CO emission
measurements (e.g., Sanders, Scoville, \& Soifer 1991) 
as well as NICMOS-2 images of their inner regions (Scoville 
et al. 2000; Alonso-Herrero et al. 2000; Alonso-Herrero et al. 2001).  
Thus, we are able to choose galaxies with very compact nuclear
CO associated with high molecular gas mass and with simple nuclear structure
(usually an $r^{1/4}$ or exponential profile).

Near-infrared images ($H$ and $K_s$) of the majority of galaxies in the LIRG 
sample were obtained with 
the wide-field PISCES camera (McCarthy et al. 2001) at the 2.3\,m Bok and 
6.5\,m MMT telescopes.  Table 1 
gives the dates and exposures times for each galaxy in both bands.  All images
were corrected for quadrant cross-talk effects known to be present in
$1024\times1024$ HAWAII arrays, using a custom C program written by Roelof 
de Jong.  The images were then dark-subtracted
with a combination of 11 dark frames of the appropriate exposure time.  Each
frame was divided by a normalized flat-field created by a median combination 
and sigma-clip rejection of the dithered science
frames.  Bad pixels were masked out using a script that substitutes known
hot or dead pixels with the value of the mode of the image.  The PISCES 
field-of-view is an inscribed circle, so the outer edges of the images were 
also masked.  Frames were corrected for geometric distortion, although
PISCES' distortion is known to be less than a 3\% effect at the field edge 
and so does not significantly contribute to poor image 
quality or errors in photometry.  Images were then aligned and stacked using 
standard IRAF tasks.  Seeing was $\sim$\,1$\farcs$2-1$\farcs$5 for the
images taken at the 2.3\,m and $\sim$\,0$\farcs$6 for images taken at the 
MMT. For two of the galaxies, NGC\,1614 and IC\,883, we rely solely on the NICMOS-2
observations and do not have wide-field images.  $H$-band 
images for the entire sample are
shown in Figure 1.  

The morphologies of the galaxies vary from late-type spirals to 
irregular mergers and interactions (see Table 1).  Tidal tails and other 
features indicating recent merger activity can be seen in several of the 
LIRGs (e.g., NGC\,1614).  

Near-infrared spectra were taken at the MMT on 2000 December 5 and 
2001 March 2-3 with the FSpec spectrograph (Williams et al. 1993). 
The slit width was $\sim$\,$0\farcs7$, corresponding to two pixels. 
The seeing-limited images were smaller than the slit width for all the observations.
The 600 lines mm$^{-1}$ grating was used at $\sim$\,2.3\,$\micron$ to 
observe the first overtone CO absorption 
bandheads, where one pixel corresponds to $\sim$\,3.3\,\AA.  
Table 1 shows the exposure times for each galaxy. 
The terrestrial absorption spectrum was sampled by observing a star of type
A V - G V between each separate galaxy observation.  Cross-correlation
template stars of K5 III - M2 III were also observed.

All spectra were reduced as described by Engelbracht et al. (1998).
Frames were dark-subtracted, trimmed, masked for bad pixels, sky-subtracted, 
and flat-fielded using IRAF tasks called by a custom script.  
Multiple frames of objects were aligned and combined, and one-dimensional
spectra of objects, standard stars, and sky frames were extracted.
The one-dimensional galaxy and standard star spectra were wavelength 
calibrated using the wavelengths of OH airglow lines. Each galaxy spectrum 
was then divided by a template standard star spectrum of type K5 III - 
M2 III.  The continuum in each star and galaxy spectrum was fitted with a low-order 
polynomial that was then subtracted. The spectra are shown in Figure 2. 

The first overtone CO absorption features are one-sided, but sharp and strong 
and hence suitable for studies of dynamics.
The measured bandheads have rest wavelengths of 2.2935\,$\micron$ for 
$^{12}$CO J\,$=2\rightarrow0$ and 2.3227\,$\micron$ for 
$^{12}$CO J\,$=3\rightarrow1$ (Kleinmann \& Hall 1986).
The first CO overtones are filled in by emission by hot dust for Mrk 231. 
Therefore, for this galaxy we measured the second CO overtones at 1.67\,$\micron$,
using similar procedures for both the observations and the reductions as for
the rest of the sample.

\section{Analysis and Results} \label{sec:analysis}

\subsection{Velocity Dispersions}
To determine dynamical masses for the LIRGs, the velocity
dispersions must be calculated from the near-infrared spectra using
template stars with similar CO absorption bands.  There are three main
techniques that can be used:  the Fourier quotient method 
(Sargent et al. 1977),
the cross-correlation method (Simkin 1974; Tonry \& Davis 1979), and direct 
fitting (Franx, Illingworth, \& Heckman 1989; Rix \& White 1992).  In the 
Fourier quotient method,
the galaxy is assumed to be the convolution of a single stellar template
with a broadening function, $B$, which can be retrieved using an inverse 
Fourier transform.  The disadvantage of this method is that the errors in the
quotient are correlated, so that error analysis becomes difficult.
Also, because the broadening function is fitted in Fourier space,
absorption features from all parts of the spectrum interact with
each other, making $B$ very sensitive to template mismatches.
In the cross-correlation method, the galaxy and template are correlated
directly, and the location of the highest correlation peak is 
identified with the mean velocity.  The dispersion is estimated by subtracting
the width of the template autocorrelation peak in quadrature from the
width of the cross-correlation peak.  The drawback of this method is that
the spectra must be padded with extra zeroes, possibly with some trimming
of the spectral edges, resulting in lost information.  The direct
fitting method of Rix \& White (1992) assumes that the galaxy differs from 
the template by the
shape of the continuum and the kinematic broadening.  The broadening is
thought of as a superposition of multiple templates shifted in velocity space.
The disadvantage for this method is that numerous templates of differing
stellar types are almost
certainly needed for a good fit to the galaxy spectrum, taking up more
observing time.  Also, this type of fitting program may get trapped in
local minima if the initial guesses for radial velocity and velocity 
dispersion are not reasonably accurate.

Of these choices, we used the cross-correlation method on our
spectra.  Although the spectra have to be carefully continuum-subtracted
and padded with zeroes, we have trimmed very little information from 
either edge,
and have tested that losing a few Angstroms of spectrum does not change
the velocity dispersion results within the calculated error bars.

Velocity dispersions of the galaxies were calculated by comparing
the width of their CO absorption bandheads to those of the template standard
stars.  We first generated a series of broadened template
star spectra with Gaussian dispersions corresponding to a range in velocity 
of 20-200 km s$^{-1}$.  The broadened spectra were then cross-correlated
with the original template star spectrum, and the width of the
cross-correlation peak as a function of the Gaussian broadening was
measured.  Each galaxy spectrum was then cross-correlated with a
template star.  The width of the cross-correlation peak between the galaxy
and template star revealed the velocity broadening, based on the broadened
stellar template calibration.  Velocity dispersions and uncertainties for 
each LIRG are listed in Table 2.  Uncertainties were estimated by 
looking at the range of dispersion values obtained from comparing multiple
template stars with each galaxy.

Mkn\,231 was observed at 1.67\,$\micron$ to calculate the velocity
dispersion from the CO bandheads there.  However, due to difficulty in
subtracting the continuum at this wavelength range, we used the Tacconi et al.
(2002) value listed in Table 2 for the remainder of this work (see Shier
1995 for a discussion of continuum subtraction).

Table 2 shows a comparison of our dispersion measurements to others. 
Three of our galaxies, NGC\,1614, NGC\,2623,
and IC\,694/NGC\,3690 \footnote{There has been some confusion in the
literature regarding which galaxy is actually IC\,694; we refer readers to
Hibbard \& Yun (1999) for a discussion.}, overlap with a study 
by Shier, Rieke, \& Rieke (1996).  We find that the values for
the velocity dispersions are higher for the first two than those found by
Shier et al. (1996), while the value for NGC 3690 is in agreement.  One
possibility for the differences is the differing areas
of the galaxies sampled by us at the MMT (slit width $\sim$ $0\farcs7$) and
by Shier et al. at the Bok 2.3\,m (slit width $\sim$ $2\farcs4$). 
Where the CO band strength strongly peaks on the galaxy nucleus, our
measurements are
strongly weighted by the center of the LIRG and may overestimate the 
dispersion.  Alonso-Herrero et al. (2001) show that 
the CO photometric index in NGC\,1614 is extremely large in the
inner 0$\farcs$4 in comparison to the Pa\,$\alpha$ emission. 
Thus, an aperture effect is a likely explanation for the discrepancy
on this galaxy. For NGC\,2623, it is also possible that the difference in slit width 
is accounts for the (smaller) discrepancy in dispersion measurements,
although we do not have the benefit of high resolution NICMOS CO imaging
to provide supporting evidence.   Within the uncertainities, 
our dispersion values for IRAS 10565+2448 and 
Mkn\,273N agree with the values derived by Dasyra et al. (2006) and 
Tacconi et al. (2002), respectively.

\subsection{Light Profiles}
To obtain dynamical masses for the sample, bulge and/or disk
scale-lengths must be fitted to the large-scale near-infrared images.
Two-dimensional models of the galaxies were created using a de
Vaucouleurs, $r^{1/4}$ bulge 
and exponential disk fitting routine (Rix \& Zaritsky 1995).
The routine rebins the images onto a cylindrical grid, given a specified 
minimum and maximum radius and the coordinates of the galaxy center.  
Fed an initial guess for a set of bulge and disk parameters, the program
begins a ``biased random walk'' within a parameter box until it finds the
combination of parameters that minimize chi-square. 
Typically, a few pixels at the very centers of the galaxies are not 
included in the fit, because the fitting routine does not account for the 
seeing.  Although both $H$ and $K_s$-band images were fitted where
available, we report the results for $H$-band, where we have images 
for the whole sample.  Exponential disk and $r^{1/4}$ effective bulge 
$H$-band radii with errorbars are listed in Table 2. 

 \subsection{Mass Models}
As LIRGs are the products of mergers or are mergers in progress, it is
not clear whether the inner regions of the galaxies should be expected to
be hot spherical systems or cold disk systems.  Radial profile fits to
a large sample of LIRGs (e.g., Scoville et al. 2000) show that the best
fitting models can be $r^{1/4}$, exponential, or neither.  If the LIRGs
are ellipticals in formation, then we might expect their centers to be
kinematically hot spheroids. However, young stars with strong CO lines are formed in 
the gas that falls to the center of the galaxy and may have settled into a ring or
disk (Barnes \& Hernquist 1991).  Also, a remnant disk consisting of the old 
stellar population of an original merging spiral galaxy may still be intact.
Therefore, we calculate dynamical masses
based on both a series of spherical models and a disk model.  These models 
are also described by Shier (1995).

Since these systems may not be in dynamical 
equilibrium, one could question the use of this type of model.  Clearly, LIRGs with tidal streams and
tails are not in equilibrium in their outer parts.  However, our 
models are generated exclusively for the inner 1\,kpc of each galaxy, where
more stability might be expected and where the dynamics have been shown to
adhere well to the fundamental plane.  Very high resolution molecular line (CO)
measurements of LIRGs could be used confirm the relative 
equilibrium of the inner parts of the LIRG systems
and confirm the values of the velocity dispersion obtained from the 
2.3\,$\mu$m CO bandheads.

\subsubsection{Spherical Models}
The mass density distributions in the spherical models are taken from
Tremaine et al. (1994), who analytically derive a family of ``$\eta$-models'',
where the central density cusps are described by $\rho \propto r^{3-\eta}$,
where $\rho$ is the mass density.
All $\eta$ models have $\rho \propto r^{-4}$ as $r \rightarrow \infty$.  The
general form of the density distribution function is
\begin{equation}
\rho_{\eta} \equiv \frac{\eta}{4\pi} \frac{1}{r^{3-\eta}(1+r)^{1+\eta}},\ 0<\eta\leq3  .
\end{equation}

\noindent Models with $\eta=1$ resemble a singular isothermal sphere, 
$\eta=2$ models have $\rho\sim1/r$ at small radii, and $\eta=3$ models have 
constant density cores.  We consider only the $\eta=2$ and the $\eta=3$
models for the rest of this work, following Shier (1995).
The mass interior to the radius is given by
\begin{equation}
M_{\eta}(r) \equiv 4\pi \int _0 ^r r^2\rho_{\eta}(r) \ dr=\frac{r^{\eta}}{(1+r)^{\eta}} \ ,
\end{equation}

\noindent We combine these models with the hydrostatic equilibrium equation,
\begin{equation}
\frac{d(\rho \sigma^2_{los})}{dr} = - \rho \frac{d \Phi}{dr}
\end{equation}

\noindent where $\sigma_{los}$ is the line-of-sight velocity dispersion and
$\Phi$ is the gravitational potential, and solve for the integral form of
the velocity dispersion,
\begin{equation}
\sigma^2_{los}(r)=\frac{G}{\rho(r)} \int _r ^{\infty} \frac{M(r)\rho(r)\ dr}{r^2} \ .
\end{equation}

\noindent However, stars at all different radii in the galaxy contribute
to the observed velocity dispersion within the slit area of the spectrograph,
in essence taking an average of all of the velocities.  This average is 
weighted by the flux from each volume element.  In terms of the mass-to-light
ratio, $\gamma$, and the extinction along the line-of-sight, $e^{-\tau}$,
the observed velocity dispersion is given in terms of the line-of-sight 
dispersion by
\begin{equation}
\sigma^2_a = \frac{\int _a \sigma^2_{los}(r) \rho/\gamma e^{-\tau} dV}{\int _a \rho/\gamma e^{-\tau} dV} \ .
\end{equation}

\noindent This is a volume integral that covers the portion of the LIRG
seen through the aperture.  

We can rewrite the form of $\rho$ in terms of an 
effective or
scale radius, $R_{eff}$, that describes the physical size of the model galaxy 
and the radius at which the density power law changes.  This form is
\begin{equation}
\rho(r) = \frac{\rho_0}{(r/R_{eff})^{\eta-3}(1+r/R_{eff})^{1+\eta}} \ .
\end{equation}

\noindent We can then write the mass interior to a radius $r$ as
\begin{equation}
M(r) = \frac{R_{eff} \sigma^2_a}{G} \frac{(r/R_{eff})^{\eta}}{(1+r/R_{eff})^{\eta}} \frac{\int _ a \rho/\eta e^{-\tau}dV}{\int_a v^2_{r,\eta}(r) \rho / \eta e^{-\tau}dV} \ .
\end{equation}

\noindent In this case $v^2_{r,\eta}(r)$ is a dimensionless velocity dispersion
calculated by Tremaine et al. (1994).  The definition is given by
\begin{equation}
v^2_{r,\eta}(r) = G M_T / R_{eff} \sigma(r),
\end{equation}

\noindent where $M_T$ represents the total mass of the galaxy.
Table 3 lists the calculated LIRG masses for $\eta=2$ and 
$\eta=3$ spherical models based on our measured velocity dispersions and 
scale-lengths.

\subsubsection{Disk Models}
To create a simple disk model, we assume that the 
local velocity dispersion is small in comparison to the circular velocity,
$v_c$, so that the stars in the disk are on nearly circular orbits, given
by 
\begin{equation}
v^2_c(r) = \frac{G M (r)}{r} \ .
\end{equation}

\noindent The inclination reduces the line-of-sight velocity by $\cos i$.
Also, stars that are not on the observed major axis have components of
their motion in the plane of the sky, so that the observed velocity is
actually given by
\begin{equation}
v_{los}(r,\theta) = v_c \cos i \sin \theta \ ,
\end{equation}

\noindent where $\theta$ is the azimuthal coordinate in the galaxy.
As we have no direct information about the inclination angles for the LIRGs, 
we set $i$ equal to 30$^{\circ}$, which is the median inclination for a set 
of randomly oriented disks.  The observed velocity dispersion is the standard
deviation of the velocities of the stars within one spectrograph aperture
width.  This standard deviation is weighted by the flux received at each
point along the aperture, so that the observed velocity dispersion, 
$\sigma_a$, is given by
\begin{equation}
\sigma^2_a = \frac{\int _a \frac{GM(r)}{r} \cos^2i \sin^2\theta \rho / \gamma dA}{\int _a \rho / \gamma dA} \ .
\end{equation}

\noindent Assuming that the LIRG has an exponential
surface mass density, i.e., $\Sigma = \Sigma_0e^{r/r_{exp}}$,
the mass of the galaxy within a specific radius is
\begin{eqnarray}
M(r)=\frac{r_{exp}\sigma^2_a}{G\cos^2i} [1-e^{-r/r_{exp}}(1+r/r_{exp})] \times \nonumber \\
\frac {\int _a \Sigma/\gamma \ dA}{\int _a \frac{1-e^{-r/r_{exp}}(1+r/r_{exp})}{r/r_{exp}} \sin^2\theta\Sigma/\gamma \ dA} \ .
\end{eqnarray}

\section{Comparing Molecular Masses to Dynamical Masses} \label{sec:masses}

Table 3 shows the molecular gas masses from millimeter CO measurements
assuming the standard $^{12}$CO-to-H$_2$ conversion factor 
(Scoville et al. 2000), 
dynamical masses from other works (Shier et al. 1996; 
Alonso-Herrero et al. 2001; Tacconi et al. 2002), and the dynamical masses
from this work for the disk, $\eta=2$, and $\eta=3$ models.  All dynamical
masses are the values for the galaxy mass within the central 1\,kpc, for 
direct comparison with
the Shier et al. (1996) values.
The mass errors are calculated from the errors in the
velocity dispersions derived from the near-IR spectroscopy.  

It is interesting to compare our dynamical mass for NGC\,1614 with that from Alonso-Herrero et al. (2001).
They used Br\,$\gamma$ spectroscopy (Puxley \& Brand 1999) and HST/NICMOS
images in Pa\,$\alpha$ of an emitting ring region to calculate a dynamical
mass independent of the CO bandheads.  They assumed that the ring is 
associated with a Lindbland resonance or similar dynamical feature and that
it is close to round.  They obtained a rotational velocity for the
ring by modeling the line profile expected for a rotating ring and also 
calculated an inclination for the ring based on its observed ellipticity
($i=51^{\circ}$).  They then used the virial theorem to convert the ring's
rotational velocity to a mass.  They found a dynamical mass of 
$1.3\times10^9$\,M$_{\odot}$ within $\sim$\,620\,pc.  
They also estimated the mass in old stars as $1.2\times10^9$\,M$_{\odot}$,
and the modeled starburst mass as $0.55\times10^9$\,M$_{\odot}$, both of
which fit within our mass budget for the stellar population in that
region in addition to the Scoville et al. (2000) gas mas estimate.  They 
noted that their mass determination was lower than that of Shier et al. (1996),
and this discrepancy is increased relative to our new determination. Changes in
the estimate of the inclination angle of the system do not increase their
dynamical mass by a large enough factor to make up the difference in 
estimates.  They use the Shier et al. (1996) scale-lengths for their mass
models, but even adopting those scale-lengths with our value for the
velocity dispersion does not bring the estimates into agreement. This
comparison emphasizes the sources of error in the dynamical mass determinations.
It suggests that, at least for some of our sample, our measurements should
be interpreted as upper limits because the dispersions are measured
only on the nuclei and a compact nuclear star cluster with very strong
CO bands may bias our dispersions toward high values. Conversely, a galaxy
nucleus dominated in K-band by an active nucleus would have a bias in
dispersion measurements toward anomalously small values; however, except
for Mkn\,231, none of our sample have strong AGN. 

In the cases of 
NGC\,1614 and NGC\,2623, the higher dynamical mass estimates now allow for the 
molecular gas portion calculated by Scoville et al. (2000).  The most
favorable (largest dynamical mass estimates) cases 
leave $\sim3\times10^9$\,M$_{\odot}$ for the stellar population 
in NGC\,1614 and $\sim7\times10^9$\,M$_{\odot}$ for NGC\,2623.  
However, for half (5) 
of galaxies in the sample, we find total dynamical masses that
are lower than the molecular gas masses predicted from millimeter CO 
measurements using the standard conversion between CO and H$_2$.  The gas
masses range from $\sim$\,1.5 to 10 times the dynamical masses in these
LIRGs.  This
is not reasonable, assuming that at least some of the mass in the inner
regions of these galaxies is in the form of stars.  For two other 
galaxies, NGC\,3690 and Mkn\,273, almost the entire dynamical mass budgets 
would have to be taken up by molecular gas for the estimates to agree.

The results for these ten LIRGs and ULIRGs have strong implications for 
theories on the efficiency
and mechanisms for star formation in extreme environments such as the
inner portions of LIRGs.  The Schmidt law (Schmidt 1959) is widely
applied to fit star forming behavior, where the star formation rate
is given by a gas density power law,
\begin{equation}
\Sigma_{SFR} = A \Sigma^N_{gas} \ , 
\end{equation}  

\noindent where $A$ is the absolute star formation rate efficiency.  
Observations generally place the value of $N$ in the range 1-2, depending
on which star formation tracers are used.  Kennicutt (1998) explored the
changes in $A$ and $N$ in the high density environments of a sample of 
luminous infrared starburst galaxies, including several LIRGs in our
sample.  He found a well-defined Schmidt law for the starbursts, with
$N=1.3-1.4$, and calculated that the global star formation efficiencies
are larger than for a sample of normal spirals by a median factor of 6.
However, he cautioned that the CO/H$_2$ standard conversion factor led to
scatter in correlations between the star formation rate and molecular gas
density.  For instance, it may be that the conversion factor is valid in
regions with solar metallicity but that it underestimates the H$_2$ mass
in metal-poor galaxies.  He found that metal-rich spirals showed a better
defined star formation rate versus molecular hydrogen gas density, in support
of the fact that the CO/H$_2$ conversion might indeed be wrong for some
environments.  He further
showed that lowering gas masses in the starburst galaxies by a factor of
two, corresponding to a conversion factor half of the standard value, 
increased $N$ from 1.4 to 1.5. With the indications that
the overestimate of the conversion factor may be 
more like a factor of $3 - 4$ than two (this work; 
Maloney \& Black 1988; Solomon et al. 1997; Downes \& Solomon 1998; Lisenfeld,
Isaak, \& Hills 2000), it seems likely that $N$ lies between 1.5 and 2. 
As a result, the star formation efficiency must increase substantially
with increasingly luminous star formation.

Our results may also indicate what mechanisms control the star formation
in merging starbursts.  There
are three basic physical processes behind intense starbursts such as those
in LIRG environments:  gravitational instabilities
in nuclear gas disks (Kennicutt 1998), efficient cloud-cloud collisions 
driven by gas motion in their central regions which are proportional to
a power of local volumic density, $\propto \rho^2$ (Scoville et al. 1991), and 
dynamical disturbances or tidal forces, perhaps, for example, from 
supermassive binary black holes
(Taniguchi \& Wada 1996).  Schmidt laws with $N\sim\,1-1.5$ favor the
gravitational instability scenario, while laws with $N\sim\,2$, such as
are suggested for our LIRGs, favor star
formation triggered by cloud-cloud collisions.  The third mechanism of
dynamical disturbances also relies on gravitational instability formalism, so 
is generally associated with lower values of $N$.  

Taniguchi \& Ohyama (1998)
found that, for two samples of high luminosity starburst galaxies with 
millimeter CO measurements, the surface density of the far-infrared
luminosity was proportional to the derived H$_2$ surface mass density, with
$N\sim\,1$, implying that the most likely star formation mechanism is
gravitational instability in nuclear gas disks when the standard ratio is
used.  However, Scoville, Sanders, \& Clemens (1986) suggested that cloud-cloud
collisions account for the high rates of star formation in LIRGs, with
this process being the dominant mode for the formation of high-mass stars.
They explored CO data on a large sample of giant molecular clouds and
suggested that any model attempting to mimick massive star formation must
account for both the formation of stars deep in the interior of clouds
and the lower efficiency in high-mass clouds.  A model that accounted
for both of these factors was one in which the star formation is triggered
by cloud-cloud collisions, so that OB star clusters appear on the edge
of and near the center of merged clouds.  The lower gas masses implied by
the calculated dynamical masses for our sample suggest larger
values of $N$, which can only be supported by the more efficient, high mass 
star formation allowed by cloud-cloud collisions.

Based on our dynamical masses and the works cited above, we conclude that 
at least some
LIRGs do not harbor the enormous molecular gas masses
calculated from CO emission measurements and that the standard conversion 
factor from CO to H$_2$ derived from local giant molecular clouds cannot be 
used in the dense inner environments of starburst galaxies.  Our dynamical
masses essentially place upper limits on the amount of molecular gas likely
to be at the centers of LIRGs.  We further
predict that the star formation efficiencies are higher than those derived from
the standard conversion values and that the dominant source of star formation
in LIRGs is likely to be cloud-cloud collisions.  We expect that
LIRGs do not have prohibitively high extinctions at near-infrared
wavelengths and anticipate that much of their structure and starburst 
information is well diagnosed with $H$ and $K_s$-band observations.  This
conclusion is supported by mid-infrared observations such as 
Gallais et al. (2004) where A$_V$ is found to be on the order of tens 
rather than hundreds.

\section{Summary}  \label{sec:summary}

We have explored the dynamics of a sample of ten
luminous and ultraluminous infrared galaxies using near-infrared imaging and
spectroscopy.  Velocity dispersions for the galaxies were calculated from
the CO absorption bandheads present at $2.3\micron$.  Exponential disk and 
effective bulge
scale-lengths for each LIRG were derived using two-dimensional model fits
to $H$-band images.  These results were used in conjunction with a series
of spherical and disk-like density profile models to determine total
dynamical masses for the galaxies.

The dynamical masses for the LIRGs are in the range 
$\sim\,2\times10^9-4\times10^{10}$\,M$_{\odot}$ for the innermost 1\,kpc.
We compare these masses to estimates of molecular gas mass derived from
interferometric millimeter wavelength observations of CO (e.g., Bryant
\& Scoville 1999).  The molecular gas masses are usually calculated
assuming a standard conversion between CO luminosity and the amount of
H$_2$, based on ratios found for giant molecular clouds in the Milky Way.
This conversion has come under some criticism from both theoretical (Maloney
\& Black 1988) and observational (Shier, Rieke, \& Rieke 1996) points
of view, and it has been suggested that using such a conversion in the
extreme environments of the nuclei of starburst galaxies overestimates
the amount of molecular hydrogen present.  In agreement with these 
criticisms, we find that for over half of the LIRGs in our sample, the 
molecular gas masses exceed or fill the entire dynamical mass budget,
an unreasonable result considering the stellar luminosity that is
observed.  We conclude that, at least for some LIRGs, the standard conversion
is an inappropriate tool for deriving molecular gas masses.

These results have wide implications for the efficiency of star formation
in LIRGs and for the physical mechanisms responsible for such star formation.
Decreased amounts of molecular gas, as implied by our dynamical masses,
allow an increase in the power of the global Schmidt law, suggesting
that the efficiency and rate of star formation in LIRGs is higher than
predicted by millimeter observations alone (Kennicutt 1998).  Such large 
star formation rates are thought to have been created by efficient 
cloud-cloud collisions in the inner regions of the LIRGs, dependent on
the square of the density in the area (Scoville, Sanders, \& Clemens 1986),
rather than by gravitational instabilities in nuclear disks or other
dynamical or tidal disturbances (Taniguchi \& Ohyama 1998).  We also
conclude that the huge amounts of extinction implied by large molecular gas
masses may not be present and that near-infrared observations are often 
adequate probes of the dynamics and starburst activity associated with LIRGs.

\acknowledgments
J. L. H. thanks Chad Engelbracht for advice on reducing the near-infrared
spectra, and Almudena Alonso-Herrero and Emeric Le Floc'h
for comments on early drafts of this
manuscript.  We also thank Don McCarthy for supporting the PISCES observing
runs, and Rose Finn for obtaining near-infrared images for two of the
LIRGs.  This research made use of the NASA/IPAC Extragalactic Database (NED) 
which is operated by the Jet Propulsion Laboratory, California Institute of 
Technology, under contract with the National Aeronautics and Space 
Administration.

\clearpage
\begin{deluxetable}{lcccccccc}
\tabletypesize{\tiny}
\tablecaption{The LIRG Sample and Observations}
\tablewidth{500pt}
\tablehead{
\colhead{Name} & \colhead{Morphological} & \colhead{L$_{IR}$} & \colhead{Imaging} & \colhead{Exposure} & \colhead{Imaging} & \colhead{Instrument/} & \colhead{Spectroscopy} & \colhead{Exposure}\\
\linebreak & Type & (L$_{\odot}$) & Band & Time (min) & Date & Telescope & Date & Time (min)\\
\linebreak (1) & (2) & (3) & (4) & (5) & (6) & (7) & (8) & (9)}
\startdata
NGC 1614 & B(s)c pec; H\,{\sc ii}: Sy2 & 3.8$\times 10^{11}$ \tablenotemark{a} & F160W & 3 & 1998 Feb 7 & HST/NICMOS & 2000 Dec 5 & 60\\ \tableline
NGC 2623 & merger & 3.5$\times 10^{11}$ \tablenotemark{a} & H & 21 & 2000 Dec 9 & PISCES/2.3\,m & 2000 Dec 5 & 108\\ 
\linebreak  & & & K$_s$ & 19 & 2000 Dec 9 & \\  \tableline
NGC 7674 & S(r)bc pec; H\,{\sc ii}: Sy2 & 3.1$\times 10^{11}$ \tablenotemark{a} & H & 21 & 2000 Dec 9 & PISCES/2.3\,m & 2000 Dec 5 & 48\\
\linebreak & & & K$_s$ & 20 & 2000 Dec 9 & \\  \tableline
IRAS 10565+2448	& H\,{\sc ii} & $10^{12}$ \tablenotemark{a} & H & 13 & 2001 May 8 & PISCES/2.3\,m & 2000 Dec 5 & 36\\
\linebreak & & & K$_s$ & 8 & 2001 May 8 & & 2001 Mar 2 & 72\\  \tableline
IRAS 17208-0014 & Sbrst H\,{\sc ii} & 2.5$\times 10^{12}$ \tablenotemark{a} & H & 10 & 2003 Mar 13 & PISCES/6.5\,m & 2001 Mar 2 & 12\\ 
\linebreak & & & K$_s$ & 10 & 2003 Mar 13 & & 2001 Mar 3 & 36\\  \tableline
IC 694/NGC3690 & Sbrst AGN & 5.2$\times 10^{11}$ \tablenotemark{b} & H & 10 & 2001 Apr 9 & PISCES/6.5\,m & 2001 Mar 3 & 84\\ 
\linebreak & & & K$_s$ & 10 & 2001 Apr 9\\  \tableline
IC 883 & Im:pec; H\,{\sc ii} LINER & 4.0$\times 10^{11}$ \tablenotemark{c} & F160W & 6 & 1997 Nov 21 & HST/NICMOS & 2001 Mar 3 & 60\\  \tableline
VII Zw31 & H\,{\sc ii} & 8.7$\times 10^{11}$ \tablenotemark{d} & H & 7 & 2001 May 8 & PISCES/2.3\,m & 2001 Dec 5 & 144\\  \tableline
Mkn 231	& SA(rs)c? pec; Sy1 & 3.5$\times 10^{12}$ \tablenotemark{c} & H & 10 & 2001 May 8 & PISCES/2.3\,m & 2001 Mar 2 & 48\\
\linebreak & & & K$_s$ & 7 & 2001 May 8 & \\ \tableline
Mkn 273 N & Sy2 & 1.4$\times 10^{12}$ \tablenotemark{c} & H & 10 & 2003 Mar 11 & PISCES/6.5\,m & 2001 Mar 3 & 84\\
\enddata
\tablenotetext{a}{Kim et al. (1995)}
\tablenotetext{b}{Gallais et al. (2004)}
\tablenotetext{c}{Sanders et al. (1988)}
\tablenotetext{d}{Scoville et al. (1989)}
\tablecomments{Column header explanations.  Col. (1):  Galaxy name.  Col. (2):
Morphological type from the NASA Extragalactic Database or Scoville et al. 
(2000). Col. (3):  IR luminosity (L$_{\odot}$) from 8-1000\,$\micron$.
Col. (4):  Near-infrared imaging band.  
Col. (5):  Imaging exposure
time in minutes.  Col. (6):  Date of imaging observation.  Col. (7):  
Instrument and telescope used for the near-infrared imaging.  Col. (8):
Spectroscopy date of observation.  Col. (9):  Spectroscopy exposure
time in minutes.}
\end{deluxetable}

\clearpage
\begin{deluxetable}{lcccc}
\tabletypesize{\tiny}
\tablecaption{LIRG Scale-lengths and Velocity Dispersions}
\tablewidth{275pt}
\tablehead{
\colhead{Name} & \colhead{$R_{exp}$} & \colhead{$R_{eff}$} & 
\colhead{$\sigma$} & \colhead{$\sigma_{other}$} \\
\linebreak & (pc) & (pc) & (km s$^{-1}$) & (km s$^{-1}$) \\
\linebreak (1) & (2) & (3) & (4) & (5)}
\startdata
NGC 1614 & $85.8\pm58.9$ & $58.9\pm105.1$ & $164\pm8$ & $75\pm12$\tablenotemark{a}\\
NGC 2623 & $1883.0\pm399.8$ & $1007.8\pm149.2$ & $153\pm11$ & $95\pm13$\tablenotemark{a}\\ 
NGC 7674 & $4420.5\pm215.0$ & $890.8\pm181.5$ & $93\pm26$ & \nodata \\ 
IRAS 10565+2448 & $1454.3\pm643.6$ & $455.5\pm87.8$ & $141\pm4$ & $125\pm31$\tablenotemark{b} \\
IRAS 17208-0014 & $1618.6\pm25.9$ & $932.1\pm73.2$ & $125\pm28$ & \nodata \\ 
IC 694/NGC 3690 & $686.5\pm28.9$ & $12.15\pm4.19$ & $141\pm17$ & $135\pm21$\tablenotemark{a} \\ 
IC 883 & $224.6\pm18.4$ & $19.0\pm48.9$ & $151\pm5$ & \nodata \\ 
VII Zw 31 & $1533.3\pm53.4$ & $416.7\pm320.6$ & $98\pm13$ & \nodata \\ 
Mkn 231 & $573.6\pm21.1$ & $400.7\pm46.4$ & \nodata & $115\pm10$\tablenotemark{c}\\ 
Mkn 273 N & $1317.0\pm10.8$ & $732.4\pm82.0$ & $232\pm43$ & $285\pm30$\tablenotemark{c} \\ 
\enddata
\tablenotetext{a}{Shier et al. (1996)}
\tablenotetext{b}{Dasyra et al.  (2006)}
\tablenotetext{c}{Tacconi et al. (2002)}
\tablecomments{Column header explanations.  Col. (1):  Galaxy name.  
Col. (2):  $H$-band exponential disk scale-length in parsecs. 
Col. (3):  $H$-band effective radius of the bulge in parsecs.
Col. (4):  Velocity dispersion in km s$^{-1}$ from this work. 
Col. (5):  Velocity dispersion in km s$^{-1}$from other works.}
\end{deluxetable}

\clearpage
\begin{deluxetable}{lccccc}
\tabletypesize{\tiny}
\tablecaption{LIRG Masses}
\tablewidth{315pt}
\tablehead{
\colhead{Name} & \colhead{$\log M$(H$_2$)} & \colhead{$\log M_{dyn}$} & 
\colhead{$\log M_{dyn}^{disk}$} & \colhead{$\log M_{dyn}^{\eta=2}$} & 
\colhead{$\log M_{dyn}^{\eta=3}$} \\
\linebreak & (M$_{\odot}$) & (M$_{\odot}$) & (M$_{\odot}$) & (M$_{\odot}$) & (M$_{\odot}$)\\
\linebreak (1) & (2) & (3) & (4) & (5) & (6)}
\startdata
NGC 1614 & 9.68 & 9.32$\pm$0.25\tablenotemark{a} & 10.01$\pm$0.05 & 9.72$\pm$0.06 & 9.84$\pm$0.07 \\ 
NGC 2623 & 9.77 & 9.46$\pm$0.21\tablenotemark{b} & 10.00$\pm$0.06 & 10.22$\pm$0.07 & 10.11$\pm$0.07 \\ 
NGC 7674 & 10.12 & \nodata & 9.52$\pm$0.29 & 9.82$\pm$0.28 & 9.73$\pm$0.28 \\ 
IRAS 10565+2448 & 10.34 & \nodata & 10.03$\pm$0.10 & 10.16$\pm$0.10 & 10.17$\pm$0.10 \\ 
IRAS 17208-0014 & 10.71 & \nodata & 9.92$\pm$0.22 & 10.09$\pm$0.22 & 9.99$\pm$0.22 \\ 
IC 694/NGC 3690 & 10.00 & 9.76$\pm$0.24\tablenotemark{b} & 10.01$\pm$0.11 & 10.13$\pm0.11$ & 10.09$\pm0.11$ \\ 
IC 883 & 9.87 & \nodata & 10.09$\pm$0.03 & 9.22$\pm$0.03 & 9.29$\pm$0.03 \\
VII Zw 31 & 10.69 & \nodata & 9.73$\pm$0.12 & 9.60$\pm$0.12 & 9.09$\pm$0.12  \\
Mkn 231 & 10.33 & 10.60\tablenotemark{c} & 9.90$\pm$0.08 & 9.97$\pm$0.08 & 10.00$\pm$0.08 \\
Mkn 273N & 10.58 & \nodata & 10.47$\pm$0.18 & 10.62$\pm$0.18 & 10.57$\pm$0.18 \\
\enddata
\tablenotetext{a}{Alonso-Herrero et al. (2001)}
\tablenotetext{b}{Shier et al. (1996)}
\tablenotetext{c}{Tacconi et al. (2002)}
\tablecomments{Column header explanations.
Col. (1):  Galaxy name.  Col. (2): Log of the
molecular hydrogen mass from Scoville et al. (2000) in solar masses, 
assuming the standard CO-to-H$_2$ conversion factor.  
Col. (3):  Log of the dynamical mass in solar masses from other works.  
Col. (4): Log of the
calculated dynamical mass in solar masses from this work for a disk model. 
Col. (5):  Log of the calculated dynamical mass in solar masses from this work
for a spherical, $\eta=2$ model.  Col. (6):  Log of the calculated dynamical
mass is solar masses from this work for a spherical, $\eta=3$ model.}
\end{deluxetable}

\clearpage

\begin{figure}
\includegraphics[angle=-90,scale=.66]{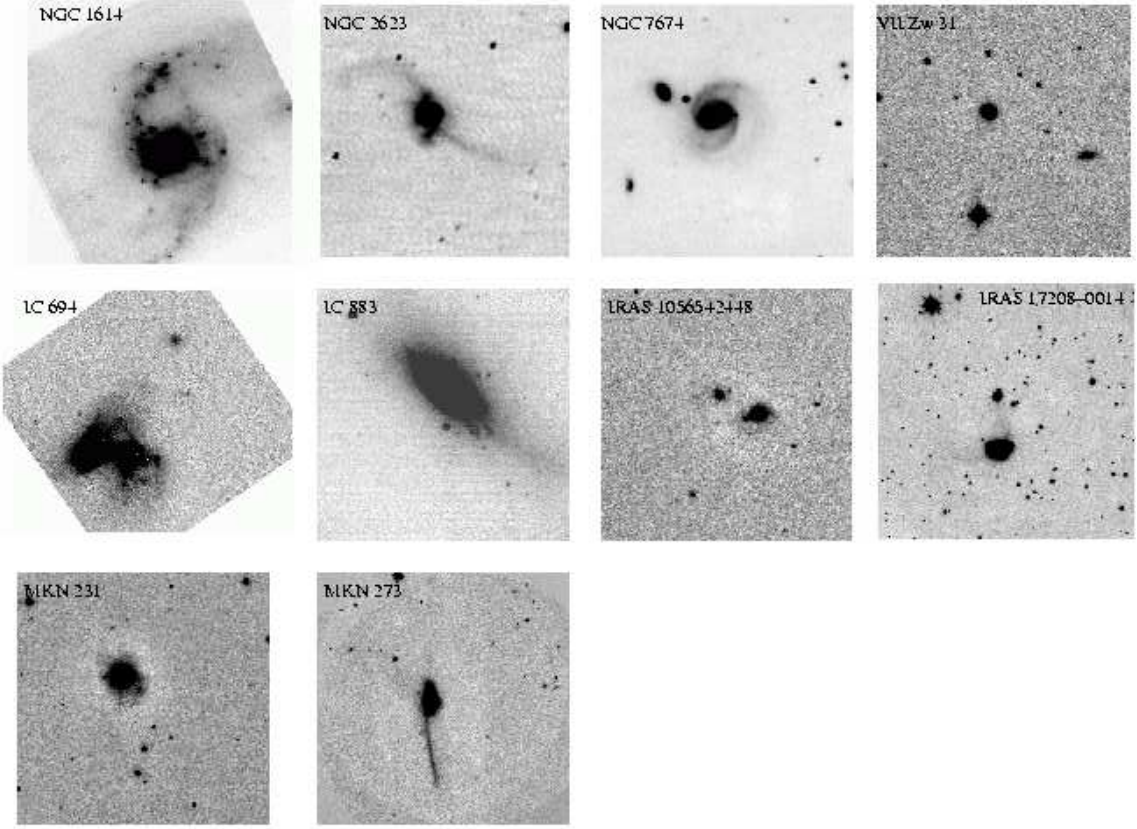}
\caption[$H$-band Images of LIRGs]{The $H$-band images of the LIRG sample
with a field of view of approximately 3$\arcmin \times 3\arcmin$, with 
north up and east to
the left.  The two HST/NICMOS images (NGC\,1614 and IC\,883) have fields of view
of $\sim$\,30$\arcsec \times 30\arcsec$.}
\end{figure}

\begin{figure}
\includegraphics[scale=0.78]{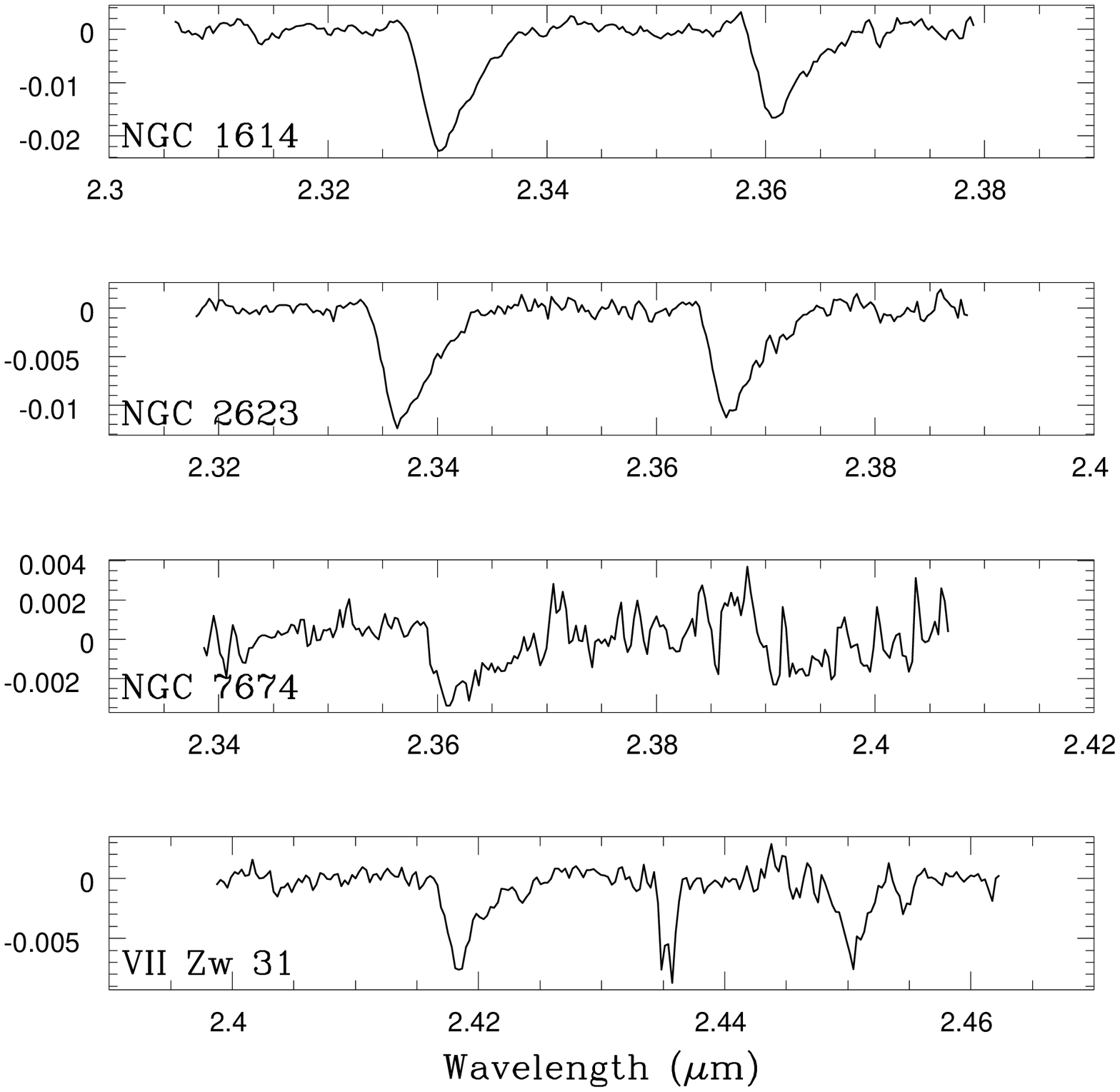}
\end{figure}
\begin{figure}
\includegraphics[scale=0.78]{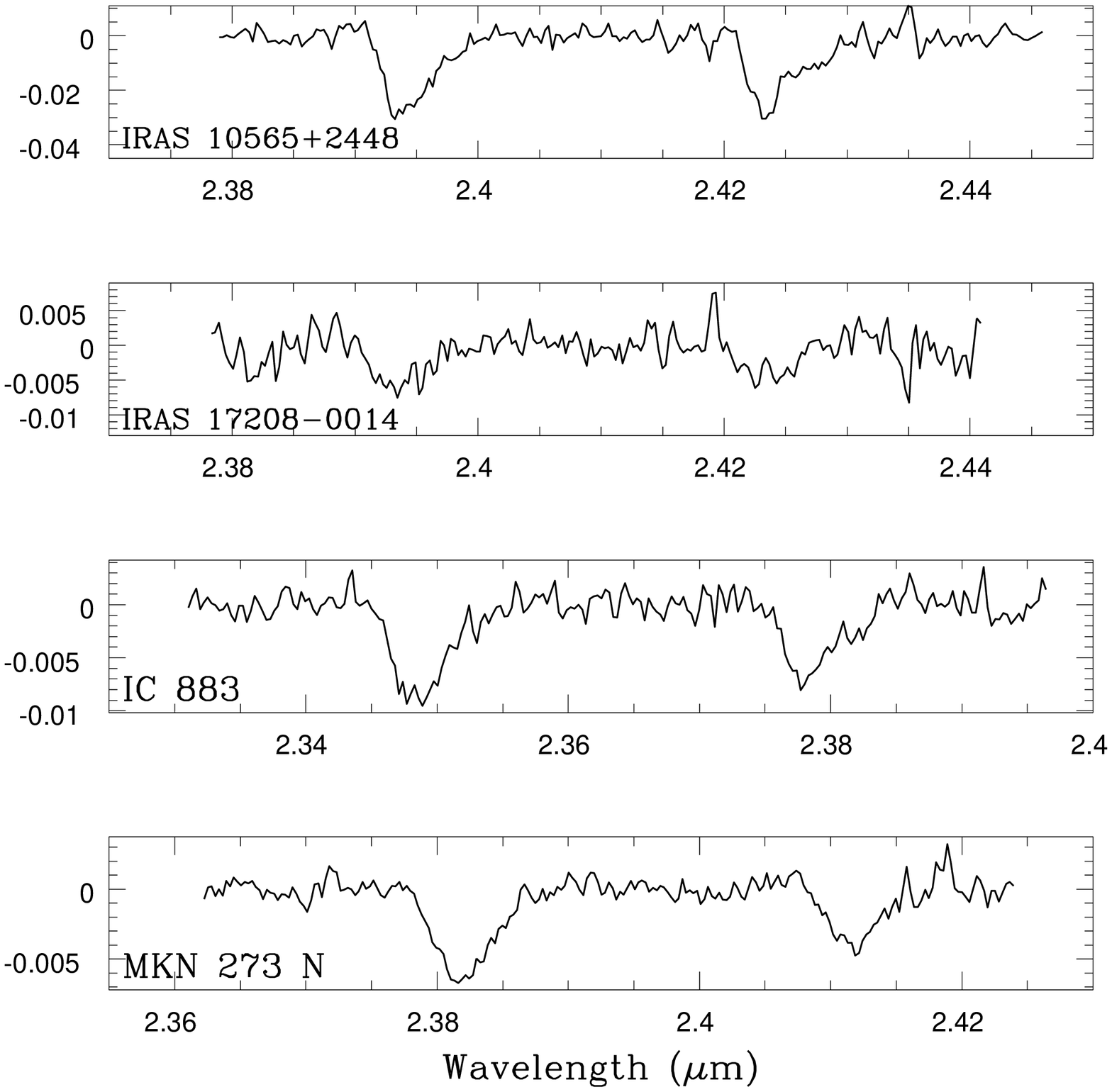}
\end{figure}
\begin{figure}
\includegraphics[scale=0.78]{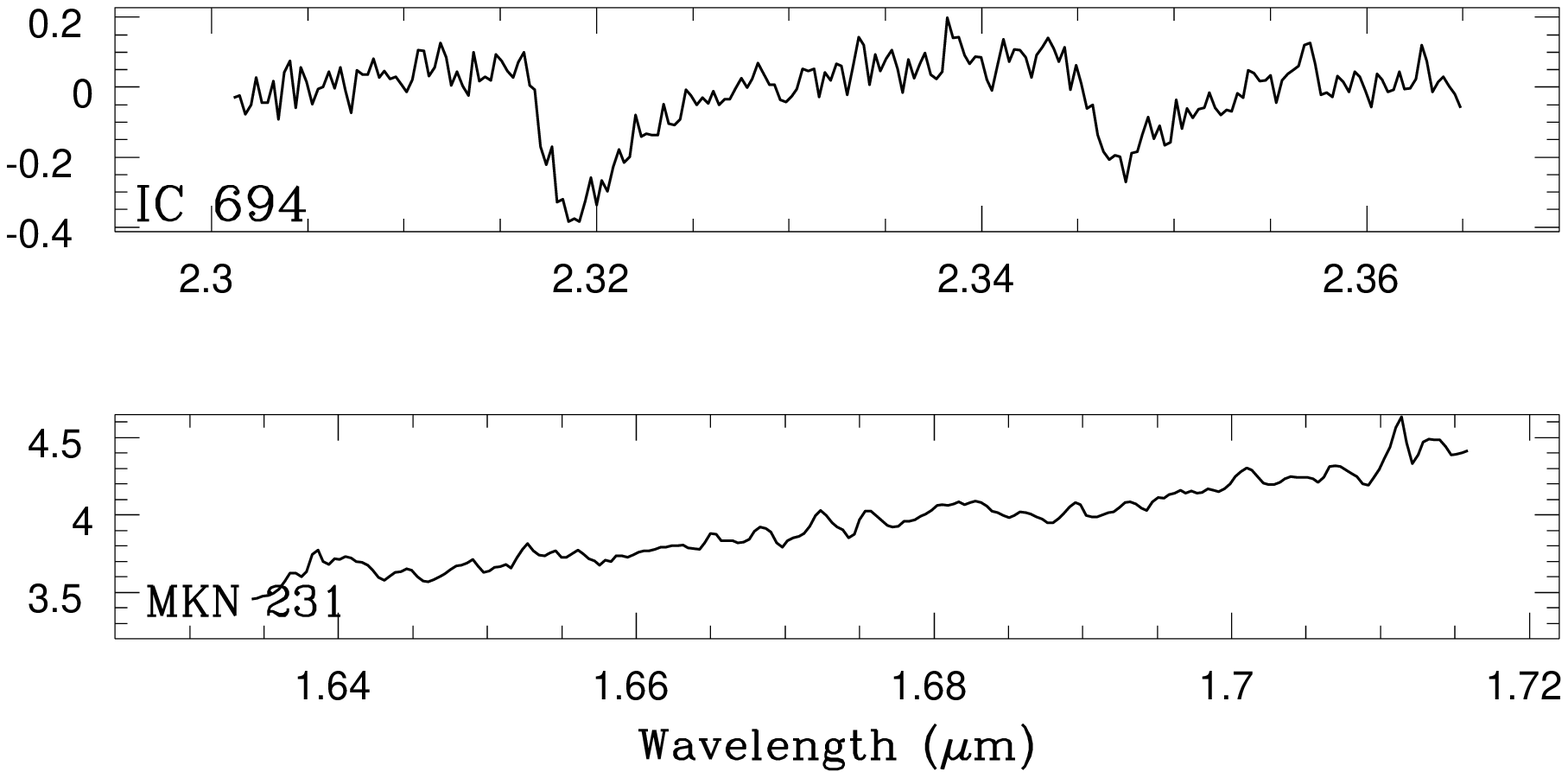}
\caption[CO Absorption Band LIRG Spectra]{One-dimensional spectra of the LIRG
sample showing the CO absorption bandheads caused by the 2-0 and 3-1 molecular
transitions.  Mkn\,231 was observed at 1.67\,$\mu$m to measure the second
overtone CO bands because of the dilution of the first overtone bands.}
\end{figure}


\begin{references}

\reference{} Alonso-Herrero,, A., Rieke, G.~H., Rieke, M.~J., \& Scoville, N.~Z.\ 2000, \apj, 532, 845 

\reference{} Alonso-Herrero, A., Engelbracht, C.~W., Rieke, M.~J., Rieke, G.~H., \& Quillen, A.~C.\ 2001, \apj, 546, 952 

\reference{} Barnes, J.~E.~\& Hernquist, L.~E.\ 1991, \apjl, 370, L65 

\reference{} Barone, L.~T., Heithausen, A., H{\"u}ttemeister, S., Fritz, T., \& Klein, U.\ 2000, \mnras, 317, 649 

\reference{} Bryant, P.~M.~\& Scoville, N.~Z.\ 1999, \aj, 117, 2632 

\reference{} Boselli, A., Lequeux, J., \& Gavazzi, G.\ 2002, \aap, 384, 33 

\reference{} Bushouse, H.~A., et al.\ 2002, \apjs, 138, 1 

\reference{} Charmandaris, V., et al.\ 2004, \apjs, 154, 142 

\reference{} Colina, L., Arribas, S., \& Monreal-Ibero, A.\ 2005, \apj, 621, 725 
\reference{} Dasyra, K.~M., et al.\ 2006, \apj, 638, 745 

\reference{} Djorgovski, S. \& Davis, M.\ 1987, \apj, 313, 59

\reference{} Downes, D. \& Solomon, P.~D.\ 1998, \apj, 507, 615

\reference{} Dressler, A., Lynden-Bell, D., Burstein, D., Davies, R.~L., Faber, S.~M., Terlevich, R., \& Wegner, G.\ 1987, \apj, 313, 42 

\reference{} Egami, E., et al.\ 2004, \apjs, 154, 130 

\reference{} Engelbracht, C. W., Rieke, M. J., Rieke, G. H., Kelly, D. M., \& Achertermann, J. M.\ 1998, \apj, 505, 639

\reference{} Farrah, D., et al.\ 2001, \mnras, 326, 1333 

\reference{} Franx, M., Illingworth, G. \& Heckman, T.\ 1989, \apj, 344, 61

\reference{} Gallais, P., Charmandaris, V., Le Floc'h, E., Mirabel, I.~F., Sauvage, M., Vigroux, L., \& Laurent, O.\ 2004, \aap, 414, 845 

\reference{} Genzel, R., Tacconi, L.~J., Rigopoulou, D., Lutz, D., \& Tecza, M.\ 2001, \apj, 563, 527 

\reference{} Hernquist, L.\ 1992, \apj, 400, 460

\reference{} Hernquist, L.\ 1993, \apj, 409, 548 

\reference{} Hibbard, J.~E., \& Yun, M.~S.\ 1999, \aj, 118, 162 

\reference{} Israel, F.~P.\ 1997, \aap, 328, 471 

\reference{} Israel, F.\ 2000, Molecular hydrogen in space, Cambridge, UK: Cambridge University Press, 2001.~xix, 326 p..~Cambridge contemporary astrophysics.~Edited by F.~Combes, and G.~Pineau des For{\^e}ts.~ISBN 0521782244, p.293, 293 

\reference{} Kennicutt, R. C., Jr.\ 1998, \apj, 498, 541

\reference{} Kim, D.-C., Sanders, D.~B., Veilleux, S., Mazzarella, J.~M., \& Soifer, B.~T.\ 1995, \apjs, 98, 129

\reference{} Kleinmann, S.~G.~\& Hall, D.~N.~B.\ 1986, \apjs, 62, 501 

\reference{} Le Floc'h, E., et al.\ 2004, \apjs, 154, 170 

\reference{} Lisenfeld, U., Isaak, K. G., \& Hills, R. 2000, \mnras, 312, 433

\reference{} Maloney, P.~\& Black, J.~H.\ 1988, \apj, 325, 389 

\reference{} McCarthy, D. W., Jr., Ge, J., Hinz, J. L., Finn, R. A., \& de Jong, R. S.\ 2001, \pasp, 113, 353

\reference{} Narayanan D., Groppi, C.~E., Kulesa, C.~A., \& Walker, C.~K.\ 2005, \apj, 630, 269 

\reference{} Puxley, P.~J.~\& Brand, P.~W.~J.~L.\ 1999, \apj, 514, 675 

\reference{} Rix, H.-W. \& White, S. D. M.\ 1992, \mnras, 254, 38

\reference{} Rix, H.-W. \& Zaritsky, D.\ 1995, \apj, 447, 82

\reference{} Sanders, D.~B., Scoville, N.~Z., \& Soifer, B.~T.\ 1991, \apj, 370, 158 

\reference{} Sanders, D. B., Soifer, B. T., Elias, J. H., Madore, B. F., Matthews, K., Neugebauer, G., \& Scoville, N. Z.\ 1988, \apj, 325, 74

\reference{} Sargent, W.~L.~W., Schechter, P.~L., Boksenberg, A., \& Shortridge, K.\ 1977, \apj, 212, 326 

\reference{} Schmidt, M.\ 1959, \apj, 129, 243

\reference{} Schweizer, F. 1986,\ Science, 231, 193

\reference{} Scoville, N.~Z.~\& Good, J.~C.\ 1989, \apj, 339, 149 

\reference{} Scoville, N.~Z., Sargent, A. I., Sanders, D. B., \& Soifer, B. T.\ 1991, \apjl, 366, L5

\reference{} Scoville, N.~Z., Sanders, D.~B., Sargent, A.~I., Soifer, B.~T., \& Tinney, C.~G.\ 1989, \apjl, 345, L25 

\reference{} Scoville, N.~Z.~et al.\ 2000, \aj, 119, 991 

\reference{} Scoville, N.~Z., Sanders, D.~B., \& Clemens, D.~P.\ 1986, \apjl, 310, L77 

\reference{} Shier, L.~M. 1995, Ph.D. Thesis, University of Arizona

\reference{} Shier, L.~M., Rieke, M.~J., \& Rieke, G.~H.\ 1994, \apjl, 433, L9

\reference{} Shier, L.~M., Rieke, M.~J., \& Rieke, G.~H.\ 1996, \apj, 470, 222 

\reference{} Simkin, S. \ 1974, \aap, 31, 129

\reference{} Solomon, P.~M., Downes, D., Radford, S.~J.~E., \& Barrett, J.~W.  1997, \apj, 478, 144

\reference{} Strong, A.~W., Moskalenko, I.~V., Reimer, O., Digel, S., \& Diehl, R.\ 2004, \aap, 422, L47 

\reference{} Tacconi, L. J., Genzel, R., Lutz, D., Rigopoulou, D., Baker, A. J., Iserlohe, C., \& Tecza, M.\ 2002, \apj, 580, 73

\reference{} Taniguchi, Y.~\& Ohyama, Y.\ 1998, \apjl, 509, L89 

\reference{} Taniguchi, Y., \& Wada, K.\ 1996, \apj, 469, 581

\reference{} Tonry, J.~\& Davis, M.\ 1979, \aj, 84, 1511 

\reference{} Toomre, A.~\& Toomre, J.\ 1972, \apj, 178, 623 

\reference{} Tremaine, S., Richstone, D.~O., Byun, Y., Dressler, A., Faber, S.~M., Grillmair, C., Kormendy, J., \& Lauer, T.~R.\ 1994, \aj, 107, 634

\reference{} Veilleux, S., Kim, D.-C., \& Sanders, D.~B.\ 2002, \apjs, 143, 315 

\reference{} Williams, D. M., Thompson, C. L., Rieke, G. H., \& Montgomery, E. F.\ 1993, Proc. SPIE, 1946, 482 

\reference{} Yamaoka, H., Kato, T., Filippenko, A.~V., van Dyk, S.~D., Yamamoto, M., Balam, D., Hornoch, K., \& Plsek, M.\ 1998, IAU Circ. 6859, 1 

\end{references}
\end{document}